\documentclass[11pt,twoside]{article}
\usepackage{asp2010}

\begin{document}

\title{3D Simulations of  Magnetic Massive Star Winds}
\author{Asif ud-Doula$^1$
\affil{$^1$Penn State W. Scranton, 120 Ridge View Dr., Dunmore, PA 18512, USA}
}

\begin{abstract}
Due to computational requirements and numerical difficulties associated with
coordinate singularity in spherical geometry, fully dynamic 3D magnetohydrodynamic (MHD)
simulations of massive star winds are not readily available.
Here we report preliminary results of the first such a 3D
simulation using $\theta^1$~Ori~C (O5.5 V) as a model.
The oblique magnetic rotator $\theta^1$~Ori~C is a source
of hard X-ray emitting plasma in its circumstellar environment.
Our numerical model can explain both
the hardness and the location of the X-ray emission from this star
confirming that magnetically confined wind shock (MCWS) is the
dominating mechanism for hard Xrays in some massive stars.

\end{abstract}

\section{Introduction}

Since the first detection by {\it Einstein} of X-ray emission from hot,
luminous stars \citep{Har1979},
a favored scenario has been that these X-rays
arise from embedded shocks generated by intrinsic instabilities in the
radiatively driven stellar winds of such stars
\citep{LucWhi1980,Owo1988,Fel1997}.
For several O-type supergiants (e.g. Zeta Pup, Zeta Ori), the
relatively soft, broadened emission line spectrum observed by
{\it Chandra} seem generally to support this paradigm.
However, in several other hot stars (e.g. $\tau$~Sco,
$\theta^{1}$~Ori~C, $\sigma$~Ori~E), the
much harder, narrow emission-line spectrum require a different explanation.
Shock velocities needed to produce ca. 1 keV -ray in these stars
are of order 1000 km/s or so. Such velocity contrasts are difficult to
achieve in line-driven instability scenario.

A more plausible scenario is represented
by a {\it Magnetically Confined Wind Shock} (MCWS) model,
in which stellar wind upflow from opposite footpoints of closed magnetic
loops collides to form strong, stationary shocks near the loop apex
\citep{BabMon1997b,BabMon1997a}. This model has been quite successful in
explaining hard x-rays in Ap and Bp stars. However,
\citet{Gag2005} have shown that 2D fully dynamic magnetohydrodynamical (MHD)
model can also explain both
the hardness and the intensity of X-ray emission from $\theta^{1}$~Ori~C
which is an O5.5 star.

Over the past decade, such 2D MHD models have been applied  toward interpreting
Chandra X-ray spectra of hot stars in general, which have resulted in a quite extensive
series of papers \citep[see][]{udDOwo2002,udD2003,OwoudD2004,udD2006,udD2008,udD2009}.
Although such axisymmetric 2D models work well for stars that are slow rotators,
they fail to capture more dynamic winds of oblique rotators which naturally
have - especially when rotation is rapid -
lateral structures and require full 3D calculations.
Here, we present one of the first such fully dynamic 3D MHD simulation
of a massive star wind, $\theta^{1}$~Ori~C.
Due to its slow rotation, we are able to model
its wind as a field-aligned rotator. As we shall see below, even such
a slow rotator develop lateral structures that otherwise are absent in 2D models.

\section{{\it Chandra} X-ray Observations of $\theta^1$~Ori~C}
{\it Chandra} grating observations performed by \citet{Gag2005},
have shown that in $\theta^1$~Ori~C most of the x-ray emitting
plasma is very hot and can be fitted by a global temperature
of ca. 30 MK. Surprisingly, this temperature does not
vary with phase. Moreover, the width of spectral lines
are very narrow of the order few hundred km/s.

Both of these properties are well explained by our fully dynamic 2D MHD simulations
\citep{Gag2005}. However, the analysis of {\it Chandra} data shows that
most of the hot gas is located around $1.5 R_\ast$ while our simulations
place the hot gas at $\sim 2.0 R_\ast$. One of the shortcomings of these 2D simulations
is that they impose artificially axisymmetry which does not allow for
complex structures seen in the simulations to break in azimuthal
directions which in principle could lead to softer
x-ray originating closer to the stellar surface.
Here we attempt to investigate such a possibility by extending our
2D simulations to full 3D.

\section{3D MHD Model}
We study the dynamical competition between field
and wind by evolving our 3D MHD simulation from an initial condition
at time $t=0$, when a dipole  magnetic field and field-aligned rotation is suddenly
introduced into a previously relaxed, 1D spherically symmetric CAK wind.
Much of the numerical procedures and stellar parameters used are described in
\citet{udDOwo2002} and \citet{Gag2005}.
However, the work presented here in addition
includes full energy equations along with radiative cooling
and a moderate stellar rotation of 10 km/s.
We perform our calculation on a 300x90x90 grid using publicly available
Zeus-MP code, which is the parallel version of Zeus-3D \citep{StoNor1992}.
We use spherical geometry.

\begin{figure}
\plottwo{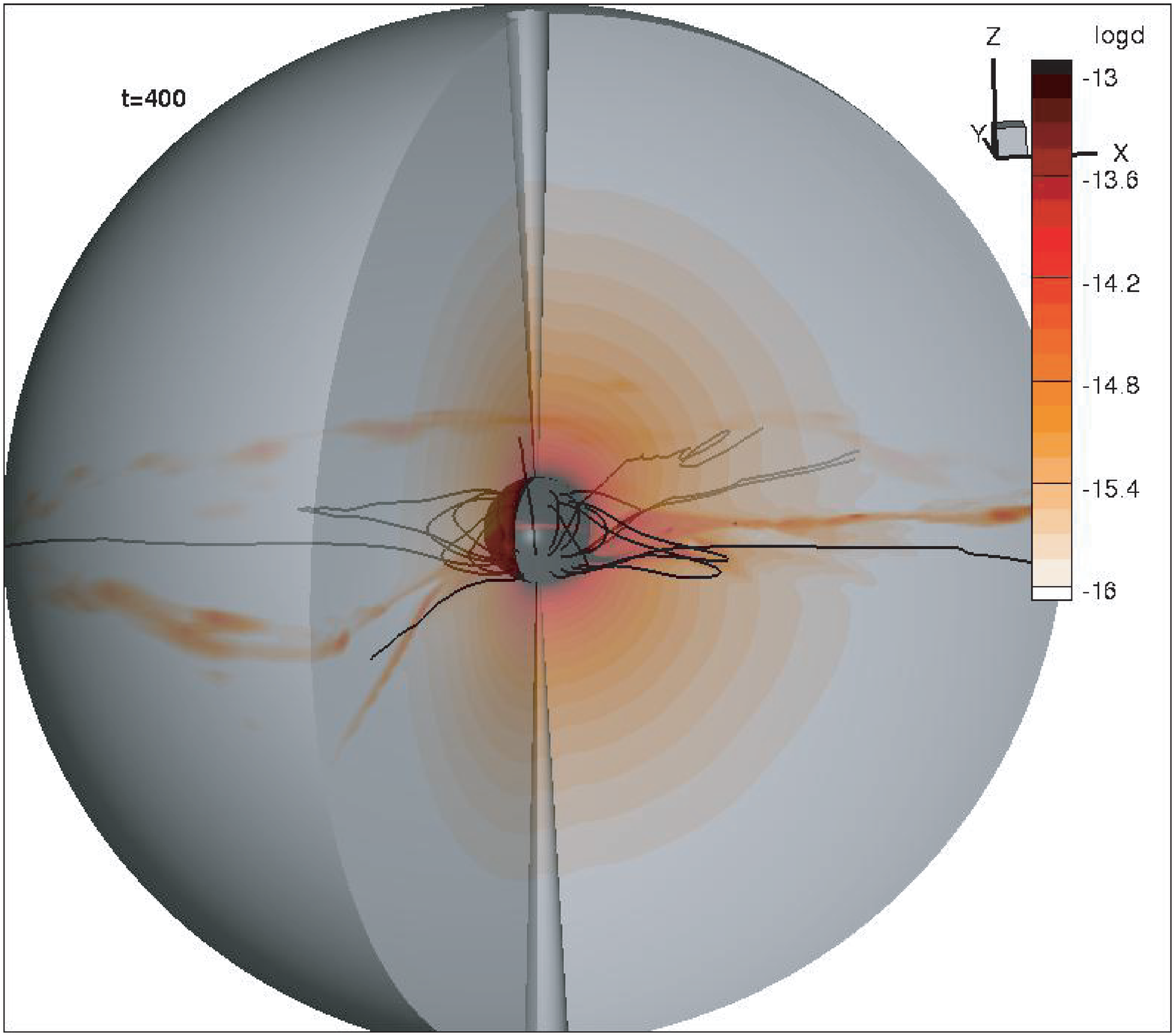}{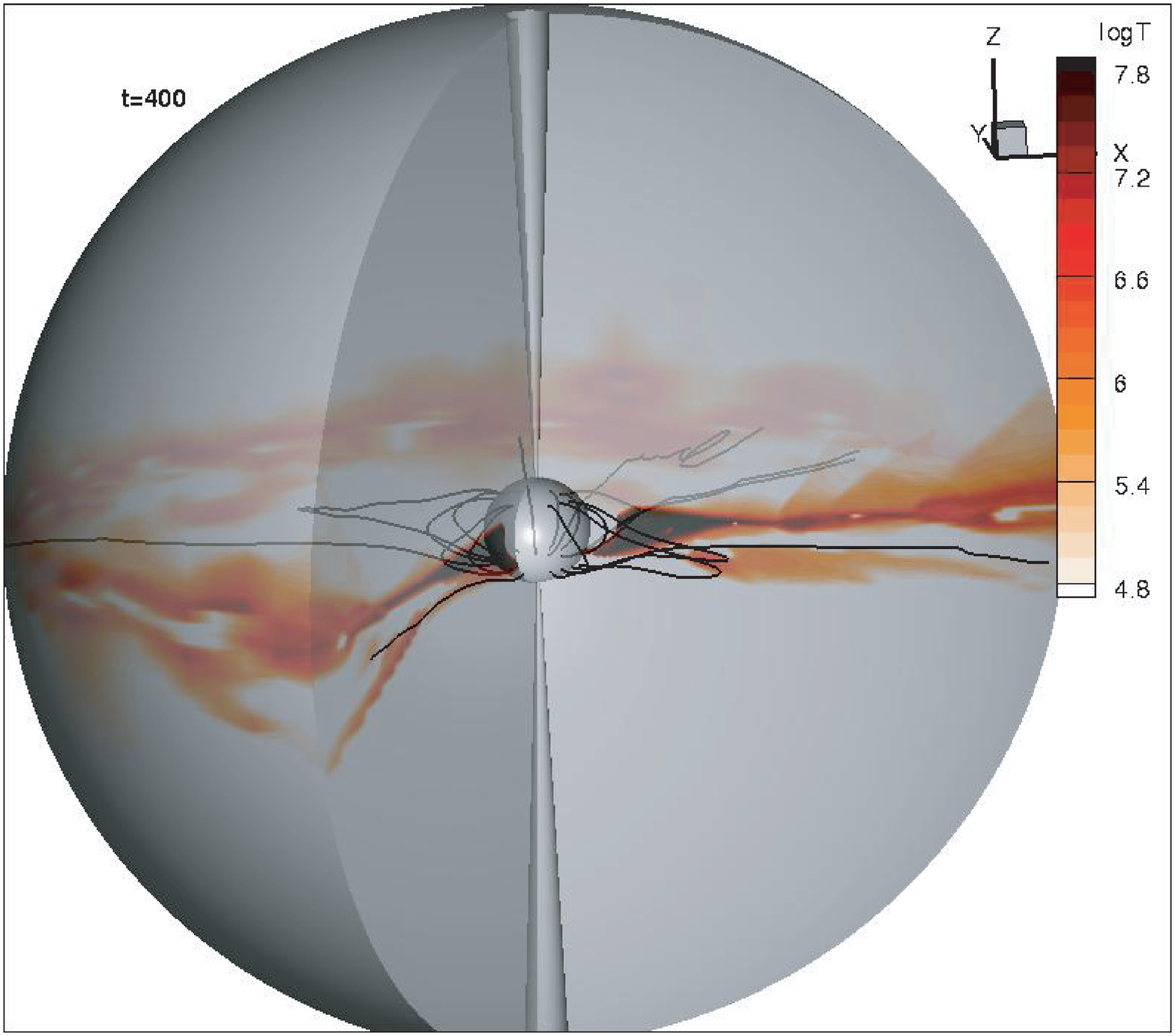}
\epsscale{0.23}
  \caption{Snapshot of logarithm of density (left panel) and temperature (right panel)
  taken at time t=400 ks. Note lateral structures are not axisymmetric.
  Black lines represent the magnetic field lines.}
\label{fig1}
\end{figure}

\begin{figure}

\plottwo{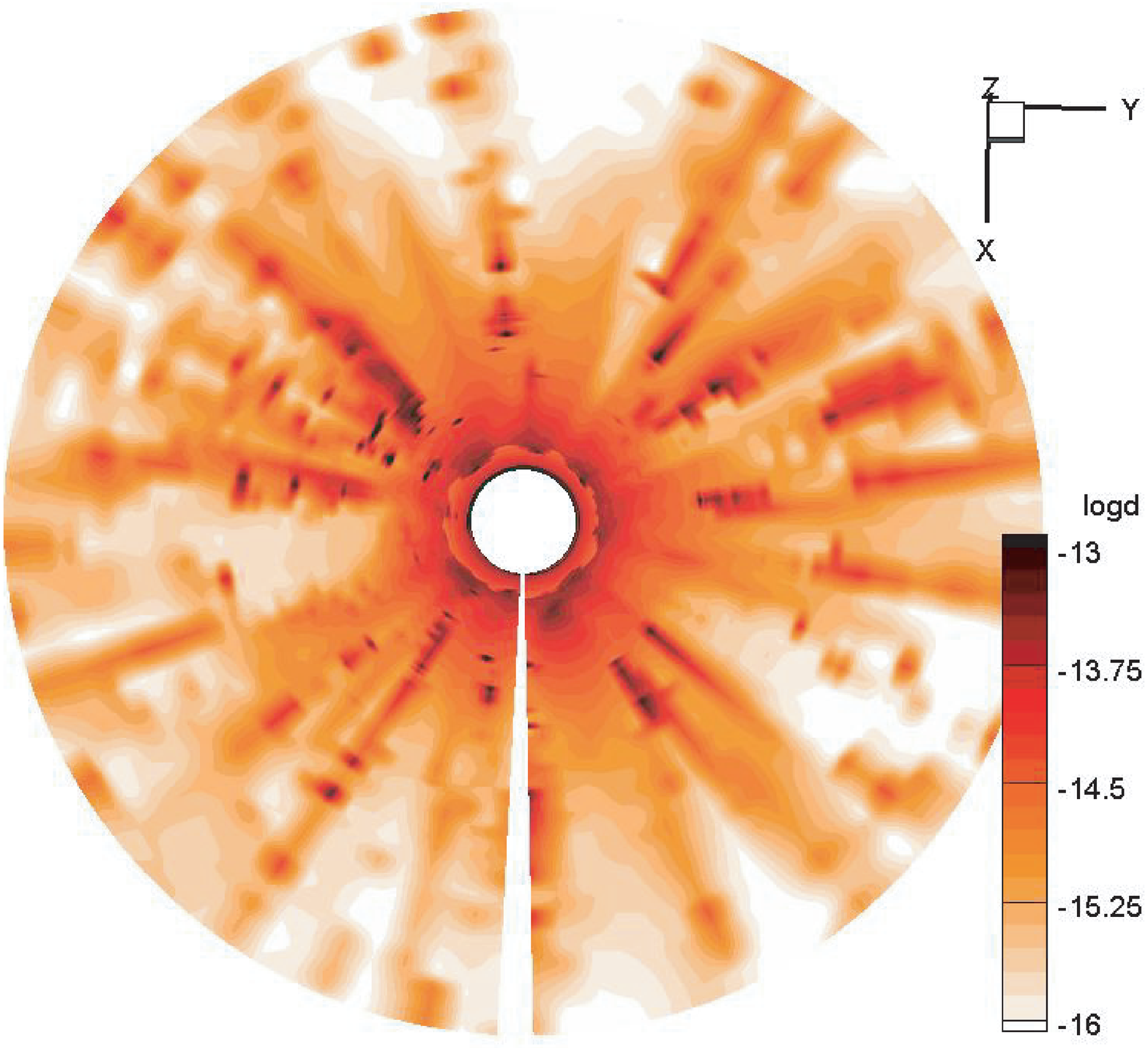}{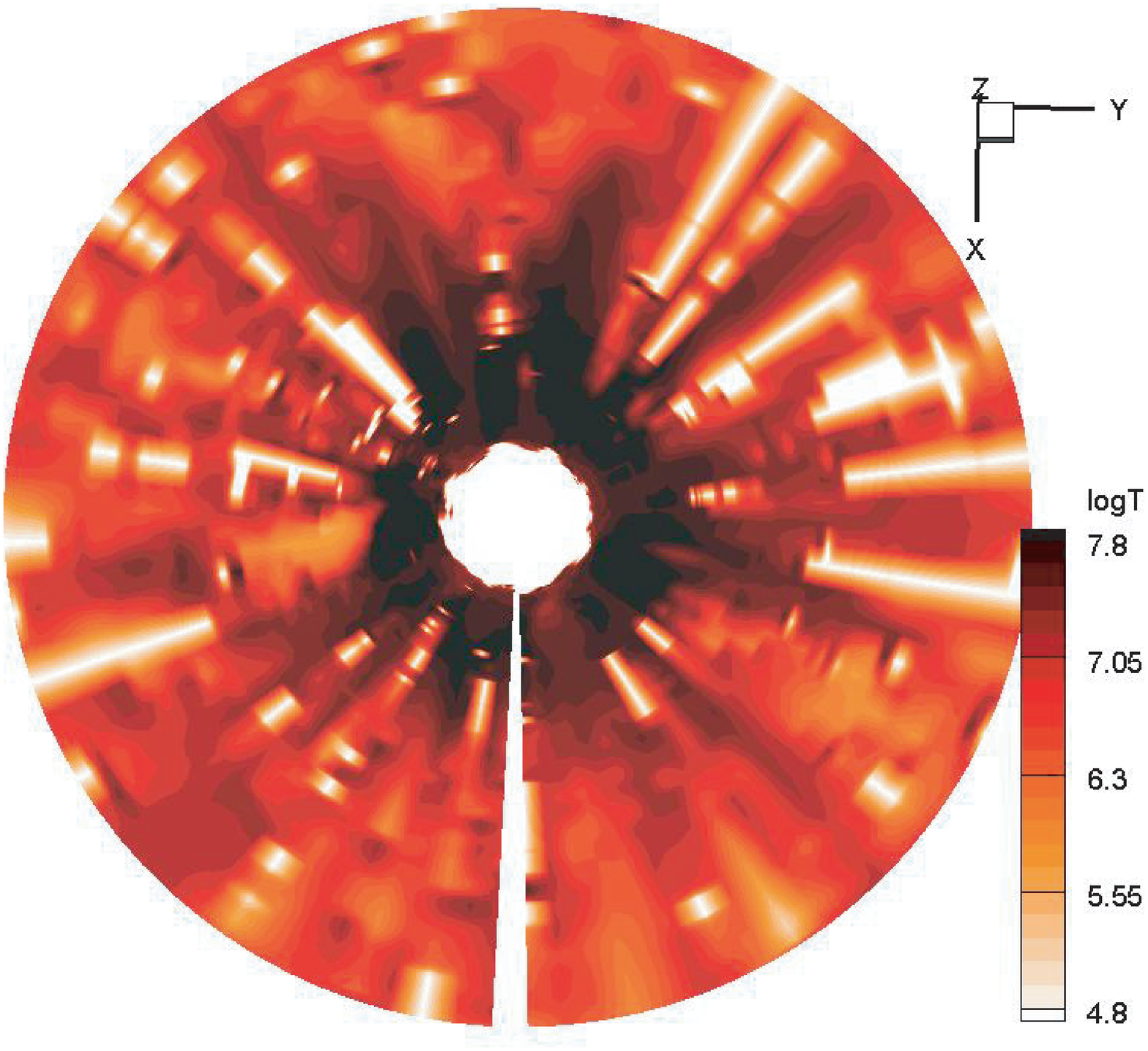}
\epsscale{0.30}
  \caption{Slice through the magnetic equator at time t=400 ks showing logarithm of density
  (left panel) and temperature (right panel.}
\label{fig2}
\end{figure}
\subsection{Dynamical Evolution}

This artificial, sudden
introduction of two dynamically incompatible states - static
dipole field and outflowing wind - leads in time to formation
of certain transient discontinuities in field and flow,
e.g., the pointed kinks in field. But these quite
quickly propagate away within a dynamical timescale of about 25 ks,
with the net overall effect of
stretching the initial dipole field outward, opening up closed
magnetic loops and eventually forcing the field in the outer
wind into a nearly radial orientation. Just like in
2D models, within the closed loop regions, wind from opposing hemispheres
collide along the magnetic equator, shock and eventually cool
creating dense equatorial structures.  Outside
and well above the closed region, the flow is quasi steady,
although now with substantial channeling of material
from higher latitudes toward the magnetic equator, with
max($v_\theta$) $>$ 500 km s$^{-1}$, even outside the closed loop. This
leads to a very strong flow compression and thus to a quite
narrow equatorial '' disk '' of dense, slow outflow.
But unlike in 2D, after about 100 ks, these structures lose their azimuthal
symmetry creating instead spike-like structures as seen in Figs. \ref{fig1}
and \ref{fig2}.

Most of these slow-moving dense structures cannot be driven by radiation.
Since they are not supported by rotation either, they fall
back onto the stellar surface in a complex pattern. However, some of the material
reach velocities beyond escape speed due to momentum deposition in oblique shocks.
These dense parcels of material do escape the star and may
be responsible for X-ray flares observed on some of the massive stars.
They can also provide a mechanism for creation of
clumps, or even be responsible for discrete absorption components (DACs).

\subsection{Differential Emission Measure}

In order to understand better the nature of  x-ray in our model,
we calculate differential emission measure (DEM) which essentially
is a volume integration of hot gas binned in 0.1 dex of temperature.
Left panel of Fig. 3 shows the time evolution of DEM for our model. Note that
there is very little time variation in par with observations. The right panel
shows averaged over time DEM. Note that most hot gas emit
X-ray in the regime of 30 MK which again agrees quite well with
observations. Further analysis shows that these parcels
of very hot gas move very slowly, less than 200 km s$^{-1}$,
suggesting that x-ray produced by such gas will form narrow lines.

Since 3D models allows for mixing in the azimuthal direction, we expected the peak temperature
of hot gas to be lower than the one obtained from our previous 2D MHD models.
However, this does not seem to be the case with 3D model suggesting even higher temperatures,
which agree with the observations even better.
Such hot gas originate around 0.5-1.0 $R_\ast$ above the stellar surface,
as suggested by the {\it Chandra} observations.

\section{Conclusion}

Here in this work, we present one of the first fully dynamic 3D MHD simulation
of the wind of a massive star, $\theta^1$~Ori~c. Initial analysis
suggests that X-ray diagnostics from our model in terms of hardness,  the location and
time variation agree better with observations than our previous 2D MHD models.
We plan to extend our 3D MHD to oblique rotator models, and synthesize
dynamical spectrum that can than be directly compared with observations.


\begin{figure}
  \plottwo{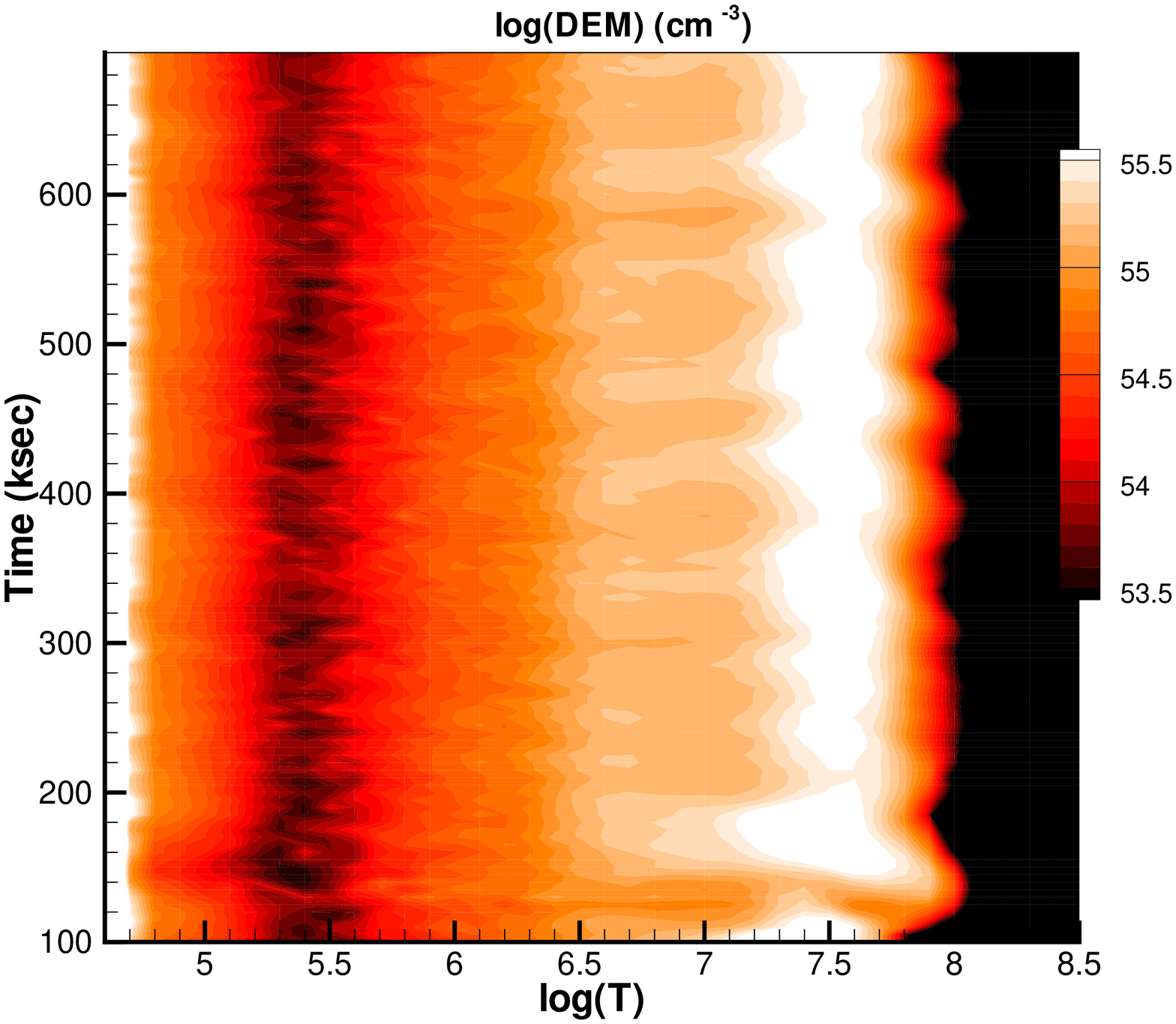}{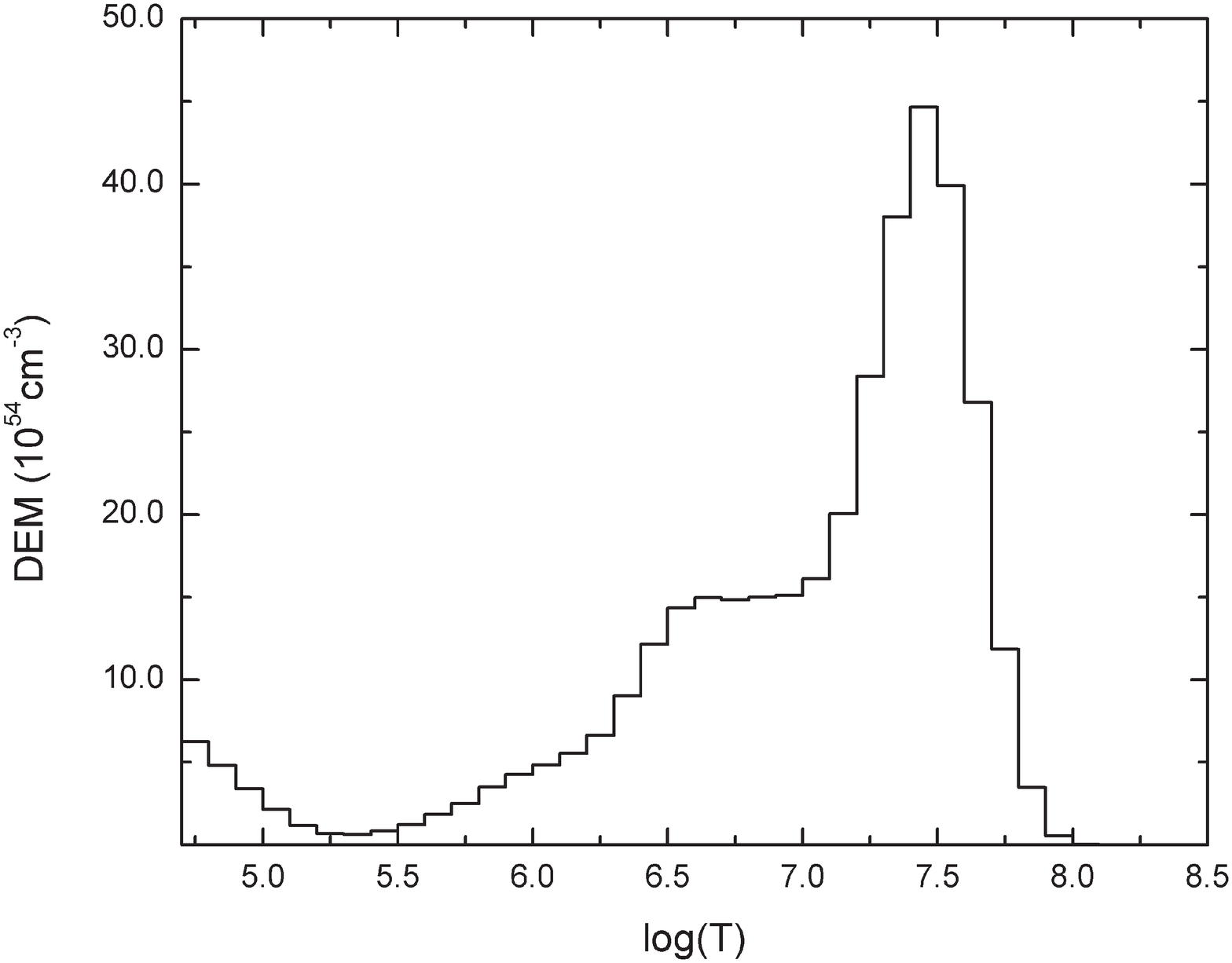}
  \epsscale{0.30}
  \caption{(a) Differential emission measure (DEM) as a function of time.
  It shows that its variation in time is relatively small.
  (b) Averaged over time DEM for our 3D MHD model. Note that
  the peak DEM temperature ca 30 MK is quite consistent with {\it Chandra} observations.}
\label{fig3}
\end{figure}


\acknowledgments
  AuD acknowledges support from NASA grant NNX11AG84G-S01.


\begin{thebibliography}{}
\expandafter\ifx\csname natexlab\endcsname\relax\def\natexlab#1{#1}\fi
\expandafter\ifx\csname url\endcsname\relax
  \def\url#1{\texttt{#1}}\fi
\expandafter\ifx\csname urlprefix\endcsname\relax\def\urlprefix{URL }\fi
\providecommand{\eprint}[2][]{\url{#2}}

\bibitem[{{Babel} \& {Montmerle}(1997{\natexlab{a}})}]{BabMon1997a}
{Babel}, J., \& {Montmerle}, T. 1997{\natexlab{a}}, \apjl, 485, 29

\bibitem[{{Babel} \& {Montmerle}(1997{\natexlab{b}})}]{BabMon1997b}
--- 1997{\natexlab{b}}, \aap, 323, 121

\bibitem[{{Feldmeier} et~al.(1997){Feldmeier}, {Puls}, \&
  {Pauldrach}}]{Fel1997}
{Feldmeier}, A., {Puls}, J., \& {Pauldrach}, A.~W.~A. 1997, \aap, 322, 878

\bibitem[{{Gagn{\'e}} et~al.(2005){Gagn{\'e}}, {Oksala}, {Cohen}, {Tonnesen},
  {ud-Doula}, {Owocki}, {Townsend}, \& {MacFarlane}}]{Gag2005}
{Gagn{\'e}}, M., {Oksala}, M.~E., {Cohen}, D.~H., {Tonnesen}, S.~K.,
  {ud-Doula}, A., {Owocki}, S.~P., {Townsend}, R.~H.~D., \& {MacFarlane}, J.~J.
  2005, \apj, 628, 986

\bibitem[{{Harnden} et~al.(1979){Harnden}, {Branduardi}, {Gorenstein},
  {Grindlay}, {Rosner}, {Topka}, {Elvis}, {Pye}, \& {Vaiana}}]{Har1979}
{Harnden}, F.~R., Jr., {Branduardi}, G., {Gorenstein}, P., {Grindlay}, J.,
  {Rosner}, R., {Topka}, K., {Elvis}, M., {Pye}, J.~P., \& {Vaiana}, G.~S.
  1979, \apjl, 234, L51

\bibitem[{{Lucy} \& {White}(1980)}]{LucWhi1980}
{Lucy}, L.~B., \& {White}, R.~L. 1980, \apj, 241, 300

\bibitem[{{Owocki} et~al.(1988){Owocki}, {Castor}, \& {Rybicki}}]{Owo1988}
{Owocki}, S.~P., {Castor}, J.~I., \& {Rybicki}, G.~B. 1988, \apj, 335, 914

\bibitem[{{Owocki} \& {ud-Doula}(2004)}]{OwoudD2004}
{Owocki}, S.~P., \& {ud-Doula}, A. 2004, \apj, 600, 1004

\bibitem[{{Stone} \& {Norman}(1992)}]{StoNor1992}
{Stone}, J.~M., \& {Norman}, M.~L. 1992, \apjs, 80, 753

\bibitem[{{ud-Doula}(2003)}]{udD2003}
{ud-Doula}, A. 2003, Ph.D. thesis, University of Delaware

\bibitem[{{ud-Doula} \& {Owocki}(2002)}]{udDOwo2002}
{ud-Doula}, A., \& {Owocki}, S.~P. 2002, \apj, 576, 413

\bibitem[{{ud-Doula} et~al.(2008){ud-Doula}, {Owocki}, \& {Townsend}}]{udD2008}
{ud-Doula}, A., {Owocki}, S.~P., \& {Townsend}, R.~H.~D. 2008, \mnras, 385, 97.
  \eprint{arXiv:0712.2780}

\bibitem[{{ud-Doula} et~al.(2009){ud-Doula}, {Owocki}, \& {Townsend}}]{udD2009}
--- 2009, \mnras, 392, 1022. \eprint{0810.4247}

\bibitem[{{ud-Doula} et~al.(2006){ud-Doula}, {Townsend}, \& {Owocki}}]{udD2006}
{ud-Doula}, A., {Townsend}, R.~H.~D., \& {Owocki}, S.~P. 2006, \apjl, 640,
  L191. \eprint{arXiv:astro-ph/0601193}

\end{thebibliography}

\end{document}